\input amstex
\input amsppt.sty

\define\X{\frak X}
\define\Y{\frak Y}
\define\pr{\operatorname{Prob}}
\define\I{\text{\bf 1}}
\define\J{\text{\bf J}}
\define\M{\Cal M}
\define\const{\operatorname{const}}
\define\tht{\thetag}
\define\pf{\operatorname{Pf}}
\define\la{\lambda}
\define\ka{\kappa}
\define\Z{\Bbb Z}
\define\R{\Bbb R}
\define\C{\Bbb C}
\redefine\P{\Bbb P}
\define\GL{\operatorname{GL}}
\define\SL{\operatorname{SL}}

\TagsOnRight
\NoBlackBoxes

\topmatter
\title
Eynard-Mehta theorem, Schur process, and their Pfaffian analogs
\endtitle
\author
Alexei Borodin and Eric M. Rains
\endauthor

\date February 11, 2005
\enddate

\abstract We give simple linear algebraic proofs of Eynard-Mehta
theorem, Okoun\-kov-Reshetikhin formula for the correlation kernel
of the Schur process, and Pfaffian analogs of these results. We
also discuss certain general properties of the spaces of all
determinantal and Pfaffian processes on a given finite set.
\endabstract
\endtopmatter

\head Introduction
\endhead

The goal of this note is to give simple proofs of Eynard-Mehta
theorem, Okoun\-kov-Resheti\-khin formula for the correlation
kernel of the Schur process, and Pfaffian analogs of these
results.

The Eynard-Mehta theorem \cite{EM} provides a determinantal formula for
marginal distributions of probability measures on $nk$-point configurations
$$
\{x_1^{(1)},\dots, x_n^{(1)}\}\cup\cdots\cup \{x_1^{(k)},\dots, x_n^{(k)}\}
$$
of the form
$$
\const \cdot \det \phi_i(x_j^{(1)}) \det W_1(x_i^{(1)},x_j^{(2)}) \cdots \det
W_{k-1}(x_i^{(k-1)},x_j^{(k)}) \det \psi_i(x_j^{(k)}).
$$
The formula was initially derived for computing the spectral
correlations of coupled random matrices, but has been used for a
number of other purposes since then. Alternative proofs of the
formula can be found in \cite{NF1}, \cite{TW2}, \cite{J2}.

The Pfaffian analog of this result gives a Pfaffian formula for
marginal distributions of probability measures of the form
$$
\const \cdot \pf\epsilon(x_i^{(1)},x_j^{(1)}) \det V_1(x_i^{(1)},x_j^{(2)})
\cdots \det V_{k-1}(x_i^{(k-1)},x_j^{(k)}) \det \xi_i(x_j^{(k)}).
$$
A variant of this formula relevant for evaluating the dynamical
correlation functions of the orthogonal-unitary and
symplectic-unitary random matrix transitions, was proved in
\cite{FNH}, \cite{NF2}.

The {\it Schur process} was introduced by Okounkov-Reshetikhin in \cite{OR}. It
is a probability measure on (generally speaking, infinite) sequences of
partitions, which in the case of finite sequences
$$
\varnothing\subset \la^{(1)}\supset \mu^{(1)}\subset \la^{(2)}\supset \mu^{(2)}
\subset \dots \supset \mu^{(T-1)}\subset \la^{(T)}\supset \varnothing
$$
takes the form
$$
\const\cdot s_{\la^{(1)}}(\rho_0^+)\,s_{\la^{(1)}/\mu^{(1)}}(\rho_1^-)
s_{\la^{(2)}/\mu^{(1)}}(\rho_1^+)\,\cdots
s_{\la^{(T)}/\mu^{(T-1)}}(\rho_{T-1}^+)\, s_{\la^{(T)}}(\rho_T^-)
$$
Here $s_\la$, $s_{\la/\mu}$ are the usual and skew Schur functions, and
$\rho_i^\pm$ are specializations of the algebra of symmetric functions. Thanks
to (Jacobi-Trudi) determinantal formulas for $s_\la, s_{\la/\mu}$, the
Eynard-Mehta theorem can be applied to evaluating the correlation functions of
the Schur process. One way of doing that is explained in \cite{J2}, although
the original derivation of the correlation functions in \cite{OR} uses
different methods. We give another way of deriving the Okounkov-Reshetikhin
formula for the correlation kernel of the Schur process from the Eynard-Mehta
theorem.

The Schur process has been used for analyzing uniformly distributed plane
partitions (or 3d Young diagrams) \cite{OR}, polynuclear growth processes
\cite{J2}, and domino tilings of the Aztec diamond \cite{J3}.

Quite similarly, using the Pfaffian analog of the Eynard-Mehta
result, we obtain the Pfaffian structure and a formula for the
correlation kernel for the {\it Pfaffian} Schur process, which
associates to the sequence of partitions above the weight
$$
\const\cdot \tau_{\la^{(1)}}(\rho_0^+)\,s_{\la^{(1)}/\mu^{(1)}}(\rho_1^-)
s_{\la^{(2)}/\mu^{(1)}}(\rho_1^+)\,\cdots
s_{\la^{(n)}/\mu^{(n-1)}}(\rho_{n-1}^+)\, s_{\la^{(n)}}(\rho_n^-).
$$
where the symmetric functions $\tau_\la$ are defined by
$\tau_\la=\sum_{\kappa'\text{ even}} s_{\la/\kappa}$. These
functions have a Pfaffian representation, see Lemma 3.1 below,
which plays a key role in the proof.

The Pfaffian Schur process was essentially introduced by
Sasamoto-Imamura \cite{SI}, with $\rho_0^+$ specializing the
symmetric functions into one variable equal to 1. They computed
the correlation functions and used them for asymptotic analysis of
polynuclear growth processes with a wall. The Pfaffian Schur
process can also be used for studying tiling models with a
symmetry condition, but further explanations of this connection go
beyond the goals of this paper.

Following our treatment of the Pfaffian Schur process, Matsumoto
in \cite{Mat2} gave a linear algebraic proof of his formulas for
the correlation functions of the shifted Schur measure, see
\cite{Mat1} for the initial derivation. The shifted Schur measures
were first introduced in \cite{TW3}.

The basic tool of our proofs is the computation of inverse of the
``Gram matrix'' of inner products for the corresponding model.
Similar ideas have been previously used in \cite{TW1}, \cite{B},
\cite{R}, \cite{J1}, \cite{J2}, \cite{J4}.

In the last section of this paper we also discuss certain general
properties of the spaces of all determinantal and Pfaffian
processes on a given finite set.

This research was partially conducted during the period one of the
authors (A.B.) served as a Clay Mathematics Institute Research
Fellow. He was also partially supported by the NSF grant
DMS-0402047.

\head 1. Eynard-Mehta theorem and its Pfaffian analog
\endhead

Let $\X$ be a finite set. A {\it random point process} on $\X$ is a probability
measure on the set $2^\X$ of all subsets of $\X$. The subsets of $\X$ will also
be called {\it point configurations}.
 Let $L$ be a $|\X|\times|\X|$
matrix whose rows and column are parameterized by points of $\X$.
For any subset $X\subset \X$ we will denote by $L_X$ the symmetric
submatrix of $L$ corresponding to $X$:
$$
L_X=\Vert L(x_i,x_j)\Vert_{x_i,x_j\in X}.
$$
If determinants of all such submatrices are nonnegative (e.g., if $L$ is
positive definite), one can define a random point process on $\X$ by
$$
\pr\{X\}=\frac{\det L_X}{\det(\I+L)}\,,\qquad X\subset \X.
$$
This process is called the {\it $L$-ensemble}.

A random point process is called {\it determinantal} if there exists a
$|\X|\times|\X|$ matrix $K$ with rows and columns parameterized by points of
$\X$ such that the {\it correlation functions}
$$
\rho(Y)=\pr\{X\in 2^\X\mid Y\subset X\},\qquad Y\subset \X,
$$
of the process have determinantal form: $\rho(Y)=\det K_Y$. The matrix $K$ is
often called the {\it correlation kernel} of the process.\footnote{Note that
the correlation kernel is not defined uniquely; conjugation of $K$ by a
diagonal matrix does not change the minors $\det K_Y$.}

\proclaim{Proposition 1.1 \cite{Ma, DVJ}} The $L$-ensemble as defined above is
a determinantal point process with the correlation kernel $K$ given by
$K=L(\I+L)^{-1}$.
\endproclaim

Take a nonempty subset $\Y$ of $\X$ and, given an $L$-ensemble on
$\X$, define a new random point process on $\Y$ by considering the
intersections of the random point configurations $X\subset \X$ of
the $L$-ensemble with $\Y$, provided that these point
configurations contain the complement $\overline \Y$ of $\Y$ in
$\X$. It is not hard to see that this new process can be defined
by
$$
\pr \{Y\}=\frac{\det L_{Y\cup\overline\Y}}{\det(\I_\Y+L)}\,,\qquad Y\subset\Y.
\tag 1.1
$$
Here $\I_\Y$ is the block matrix $\bmatrix \I&0\\0&0\endbmatrix$
where the blocks correspond to the splitting
$\X=\Y\sqcup\overline{\Y}$. We call this new process the {\it
conditional $L$-ensemble}.

\proclaim{Proposition 1.2} The conditional $L$-ensemble is a
determinantal point process with the correlation kernel given by
$$
K=\I_\Y-(\I_\Y+L)^{-1}\bigl|_{\Y\times\Y}.
$$
\endproclaim
Note that for $\Y=\X$ this statement coincides with Proposition 1.1.

\demo{Proof} Using the fact that if $B=A^{-1}$ then $\det B_X=\dfrac{\det
A_{\overline X}}{\det A}$, for any $Y\in2^\Y$ we obtain
$$
\gathered \det K_Y=\sum_{X\subset Y}
(-1)^{|X|}\det\left((\I_\Y+L)^{-1}\right)_X=\sum_{Z=\overline X\supset
\overline Y}(-1)^{|X|}\,\frac{\det(\I_\Y+L)_Z}{\det(\I_\Y+L)}\\
=\sum_{Z=\overline X\supset \overline Y}(-1)^{|X|}\,\pr\{\text{all points of
the random point configuration are in}\  Z \}\\
=\sum_{X\subset Y}(-1)^{|X|}\,\pr\{X\ \text{has no points of the random point
configuration}\}=\rho(Y)
\endgathered
$$
where the last equality is the inclusion-exclusion principle. \qed
\enddemo

Let us now state the Eynard-Mehta theorem \cite{EM}. Other proofs
of this theorem are given in \cite{NF1}, \cite{J2}, \cite{TW2}.

Consider a random point process on a disjoint union of $k$ (finite) sets
$\X^{(1)}\cup\cdots \cup\X^{(k)}$ which lives on $nk$--point configurations
with exactly $n$ points in each $\X^{(i)}$, $i=1,\dots,k$, defined by the
condition that the probability of any such point configuration equals
$$
\gathered
\pr\left\{\left\{x_i^{(1)}\right\}_{i=1}^n\cup\cdots\cup\left\{x_i^{(k)}\right\}_{i=1}^n
\right\}\qquad\qquad\qquad\qquad\qquad\qquad\qquad
 \\=\const \cdot \det_{1\le i,j\le
n}\bigl[\phi_i(x_j^{(1)})\bigr]\cdot
\det_{1\le i,j\le n}\bigl[W_1(x_i^{(1)},x_j^{(2)})\bigr]\cdots\\
\qquad\qquad\qquad\qquad\qquad \cdots
 \det_{1\le
i,j\le n}\bigl[W_{k-1}(x_i^{(k-1)},x_j^{(k)})\bigr] \det_{1\le i,j\le
n}\bigl[\psi_i(x_j^{(k)})\bigr].
\endgathered
\tag 1.2
$$
Here $\{\phi_i\}_{i=1,\dots,n}$, are some functions on $\X^{(1)}$,
$\{\psi_i\}_{i=1,\dots,n}$, are some functions on $\X^{(k)}$, and
$\{W_m\}_{m=1,\dots,k-1}$, are matrices with rows parameterized by
points of $\X^{(m)}$ and columns parameterized by points of
$\X^{(m+1)}$. The normalization constant in the right-hand side of
\tht{1.2} is chosen in such a way that the total mass of all
admissible point configurations is equal to 1. We do not address
the problem of positivity of \tht{1.2} as it does not play any
role in the sequel. It suffices to assume that the normalization
constant is finite (the total mass is nonzero).

It is convenient to organize the functions $\phi_i$ and $\psi_i$ into two
matrices $\Phi$ and $\Psi$, the rows of $\Phi$ and the columns of $\Psi$ are
parameterized by $\{1,\dots,n\}$, the columns of $\Phi$ are parameterized by
points of $\X^{(1)}$, and the rows of $\Psi$ are parameterized by points of
$\X^{(k)}$. The corresponding matrix elements are just the values of $\phi_i$
and $\psi_i$ at the corresponding points.

\proclaim{Lemma 1.3} The sum of the right-hand sides of \tht{1.2} with
``$\const$'' removed, taken over all possible point configurations is equal to
$\det M$, where
$$
M=\Phi W_1\cdots W_{k-1}\Psi. \tag 1.3
$$

Thus, $\const$ in \tht{1.2} is equal to $\det M^{-1}$, provided that $\det M\ne
0$.
\endproclaim
\demo{Proof} Follows from the well known Cauchy-Binet formula.\qed
\enddemo

In what follows we always assume that $M$ is invertible, that is
$\det M\ne 0$.

Set
$$
W_{[i,j)}=\cases W_i\cdots W_{j-1},&i<j, \\0,& i\ge j.\endcases
$$
%Let us denote by $W$ the matrix with rows and column parameterized by points of
%$\X^{(1)}\cup\cdots \cup\X^{(k)}$ whose $(i,j)$-block (i.e. values on
%$\X^{(i)}\times\X^{(j)}$) is equal to $W_{[i,j)}$.
\proclaim{Theorem 1.4 (Eynard-Mehta)} The random point process defined by
\tht{1.2} is determinantal. The $(i,j)$-block of the correlation kernel is
given by
$$
K_{ij}=W_{[i,k)}\Psi M^{-1} \Phi W_{[1,j)}-W_{[i,j)}. \tag 1.4
$$
\endproclaim
\demo{Proof} Take
$$
\X=\{1,\dots,n\}\cup\X^{(1)}\cup\cdots \cup\X^{(k)}
$$
and consider the conditional $L$-ensemble on $\X$ with $\Y=\X^{(1)}\cup\cdots
\cup\X^{(k)}$ and the matrix $L$ given in the block form by
$$
L=\bmatrix 0&\Phi&0&0&\cdots&0\\
0&0&-W_1&0&\cdots&0\\
0&0&0&-W_{2}&\cdots&0\\
\cdots&\cdots&\cdots&\cdots&\cdots&\cdots\\
0&0&0&0&\cdots&-W_{k-1}\\
\Psi&0&0&0&\cdots&0
\endbmatrix.
\tag 1.5
$$
Then this conditional $L$-ensemble is exactly the point process defined by
\tht{1.2}. Indeed, the determinant of a block matrix of type \tht{1.5} is
nonzero if and only if the sizes of all blocks are equal, and in that case the
determinant is equal to the product of determinants of the nonzero blocks up to
a sign which depends only on the size of the blocks. This observation
immediately implies that \tht{1.1} and \tht{1.2} are equivalent.

According to Proposition 1.2, in order to compute the correlation kernel we
need to invert $\I_\Y+L$.

\proclaim{Lemma 1.5} The following inversion formula for a block matrix with
square (1,1) and (2,2) blocks holds:
$$
\bmatrix A&B\\C&D\endbmatrix^{-1}=\bmatrix
-\M^{-1}&\M^{-1}BD^{-1}\\D^{-1}C\M^{-1}&
D^{-1}-D^{-1}C\M^{-1}BD^{-1}\endbmatrix, \quad \M=BD^{-1}C-A
$$
where we assume that all the needed inverses exist.
\endproclaim
\demo{Proof} The matrix in the right-hand side equals
$$
\bmatrix 1&0\\-D^{-1}C&1\endbmatrix\bmatrix
-\M^{-1}&\M^{-1}BD^{-1}\\0&D^{-1}\endbmatrix.
$$
Inverting this product we obtain
$$
\bmatrix -\M&B\\0&D\endbmatrix \bmatrix
1&0\\D^{-1}C&1\endbmatrix= \bmatrix
-\M+BD^{-1}C&B\\C&D\endbmatrix=\bmatrix A&B\\C&D\endbmatrix.\qed
$$
\enddemo

We now split $\I_{\Y}+L$ into blocks according to the splitting
$\X=\{1,\dots,n\}\cup \Y$ and use the above lemma. First of all,
$$
D^{-1}=\bmatrix 1&-W_{1}&0&\cdots&0\\
0&1&-W_{2}&\cdots&0\\
0&0&1&\cdots&0\\
\cdots&\cdots&\cdots&\cdots&\cdots\\
0&0&0&\cdots&1
\endbmatrix^{-1}=
\bmatrix 1&W_{[1,2)}&W_{[1,3)}&\cdots&W_{[1,k)}\\
0&1&W_{[2,3)}&\cdots&W_{[2,k)}\\
0&0&1&\cdots&W_{[3,k)}\\
\cdots&\cdots&\cdots&\cdots&\cdots\\
0&0&0&\cdots&1
\endbmatrix
$$

Next, $\M=BD^{-1}C-A=\Phi W_{[1,k)}\Psi$ is exactly the matrix $M$ given by
\tht{1.3}. It readily follows that $\I_{\Y}-(D^{-1}-D^{-1}CM^{-1}BD^{-1})$ is
exactly the right-hand side of \tht{1.4}.\qed
\enddemo

We now aim at proving a Pfaffian analog of Theorem 1.4. In order
to work with $2\times 2$ matrix valued matrices, we introduce two
copies of our (finite) phase space $\X$ which we will denote by
$\X'$ and $\X''$. Each point $x\in \X$ has a prototype $x'\in\X'$
and another one $x''\in\X''$.

A {\it Pfaffian $L$-ensemble} on $\X$ is a random point process on
$\X$ with probabilities of the point configurations given by
$$
\pr\{X\}=\frac{\pf L_{X}}{\pf(\J+L)}\,,\qquad X\subset\X.
$$

Here $L$ is a $|\X|\times |\X|$ skew-symmetric matrix made of
$2\times 2$ blocks with rows and columns parameterized by points
of $\X$. Alternatively, it is a $2|\X|\times 2|\X|$ matrix with
rows and column parameterized by elements of $\X'\cup\X''$. The
$2\times 2$ blocks have the form
$$
L(x,y)=\bmatrix
L(x',y')&L(x',y'')\\L(x'',y')&L(x'',y'')\endbmatrix.
$$
The matrix $\J$ is defined by
$$
\J(x,y)=\cases \bmatrix 0&1\\-1&0\endbmatrix,&x=y,\\
0,&x\ne y.\endcases
$$

A random point process is called {\it Pfaffian} if there exists a
$2\times 2$ matrix valued $|\X|\times |\X|$ skew-symmetric matrix
$K$ with rows and column parameterized by points of $\X$, such
that the correlation functions of the process have the Pfaffian
form: $\rho(Y)=\pf K_{Y}$ for any $Y\subset \X$. As in the
determinantal case, the matrix $K$ is called the {\it correlation
kernel}.

Similarly to Proposition 1.1, we have the following statement.
\proclaim{Proposition 1.6 \cite{R}} The Pfaffian $L$-ensemble as
defined above is a Pfaffian point process with the correlation
kernel $K=\J+(\J+L)^{-1}$.
\endproclaim

Once again, let us take a subset $\Y$ of $\X$ and let us consider
a new random point process on $\Y$ by taking the intersections of
the random point configuration of the Pfaffian $L$-ensemble with
$\Y$, provided that these configurations contain the complement
$\overline \Y=\X\setminus \Y$. Then the probabilities of the point
configurations for such a process are given by
$$
\pr\{Y\}=\frac{\pf L_{Y\cup\overline\Y}}{\pf(\J_{\Y}+L)}\,,\qquad
Y\subset \Y.
$$
We call this process the {\it conditional Pfaffian $L$-ensemble}.
Proposition 1.6 above is a corollary of the following more general
claim, cf. Proposition 1.2.

\proclaim{Proposition 1.7} The conditional Pfaffian $L$-ensemble
is a Pfaffian point process. Its correlation kernel is given by
$$
K=\J_{\Y}+(\J_{\Y}+L)^{-1}\bigl|_{\Y\times\Y}\,.
$$
\endproclaim
\demo{Proof} We have
$$
\pf K_Y=\sum_{X\subset Y}
\pf\left((\J_\Y+L)^{-1}\right)_X=\sum_{Z=\overline X\supset
\overline Y}(-1)^{|X|}\,\frac{\pf(\J_\Y+L)_Z}{\pf(\J_\Y+L)},
$$
and the rest is as in the proof of Proposition 1.2. Here we used
the following fact: if $A$ and $B$ are $2l\times 2l$
skew-symmetric matrices and $B=A^{-1}$ then
$$
\pf
A_{\alpha_{1},\dots,\alpha_{2m}}=(-1)^{\alpha_{1}+\dots+\alpha_{2m}}\cdot\frac{\pf
B_{\{1,\dots,2l\}\setminus\{\alpha_{1},\dots,\alpha_{2m}\}}}{\pf
B}\,. \qed
$$
\enddemo

We proceed to stating the Pfaffian analog of the Eynard-Mehta
theorem. Let us assume that our state space is a union of $k$
subsets $\X^{(1)}\cup\cdots\cup\X^{(k)}$, and consider a random
point process that lives on $2nk$ point configurations with
exactly $2n$ points in each $\X^{(i)}$, $i=1,\dots,k$. The
probability of any such point configuration is given by
$$
\gathered
\pr\left\{\left\{x_i^{(1)}\right\}_{i=1}^{2n}\cup\cdots\cup\left\{x_i^{(k)}
\right\}_{i=1}^{2n}
\right\}\qquad\qquad\qquad\qquad\qquad\qquad\qquad
 \\=\const \cdot \pf_{1\le i,j\le 2n}\bigl[\epsilon(x_i^{(1)},x_j^{(1)})\bigr]\cdot
\det_{1\le i,j\le 2n}\bigl[V_1(x_i^{(1)},x_j^{(2)})\bigr]\cdots\\
\qquad\qquad\qquad\qquad\qquad \cdots
 \det_{1\le
i,j\le 2n}\bigl[V_{k-1}(x_i^{(k-1)},x_j^{(k)})\bigr] \det_{1\le
i,j\le 2n}\bigl[\xi_i(x_j^{(k)})\bigr].
\endgathered
\tag 1.6
$$
Here $\{\xi_i\}_{i=1,\dots,2n}$, are some functions on
$\X^{(k)}$, $\{V_m\}_{m=1,\dots,k-1}$, are matrices with rows
parameterized by points of $\X^{(m)}$ and columns parameterized
by points of $\X^{(m+1)}$, and $\epsilon$ is a skew-symmetric
matrix with rows and columns parameterized by the points of
$\X^{(1)}$.

As before, it is convenient to organize $\xi_{i}$'s into one
$|\X^{(k)}|\times 2n$ matrix $\Xi$ with columns parameterized by
$1,\dots,2n$, and rows parameterized by $\X^{(k)}$; the matrix
elements are the values $\xi_i(x^{(k)})$, $x^{(k)}\in\X^{{(k)}}$.

The next statement is an analog of Lemma 1.3.

\proclaim{Lemma 1.8} The sum of the right-hand sides of \tht{1.2} with
``$\const$'' removed, taken over all possible point configurations is equal to
$\pf N$, where
$$
N=\Xi^{t}\, V_{k-1}^{t}\cdots V_{1}^{t}\,\epsilon\, V_{1}\cdots V_{k-1}\,\Xi.
$$

Thus, $\const$ in \tht{1.6} is equal to $\pf N^{-1}$, provided that $\pf N\ne
0$.
\endproclaim

Using the familiar notation
$$
V_{[i,j)}=\cases V_i\cdots V_{j-1},&i<j, \\0,& i\ge j,\endcases
$$
we have $N=\Xi^{t}\, V_{[1,k)}^{t}\,\epsilon\, V_{[1,k)}\,\Xi$.
In what follows, we will always assume that this matrix is
nondegenerate.

\proclaim{Theorem 1.9} The random point process defined by
\tht{1.6} is Pfaffian. The $2\times 2$ entries of the correlation
kernel in its $(i,j)$-block are given by
$$
\bmatrix V_{[i,k)}\Xi N^{-1}\Xi^{t}V_{[j,k)}^{t} & V_{[i,k)} \Xi N^{-1}\Xi^{t}
V_{[1,k)}^{t}\epsilon V_{[1,j)}-V_{[i,j)}\\
-V_{[1,i)}^t\epsilon V_{[1,k)}\Xi N^{-1}\Xi^t V_{[j,k)}^t+V_{[j,i)}^t&
-V_{[1,i)}^t\epsilon V_{[1,k)}\Xi N^{-1}\Xi^t V_{[1,k)}^t\epsilon
V_{[1,j)}+V_{[1,i)}^t\epsilon V_{[1,j)}
\endbmatrix
\tag 1.7
$$
\endproclaim
\demo{Proof} Take
$$
\X=\{1,\dots,2n\}\cup \X^{(1)}\cup\cdots\cup\X^{(k)}
$$
and consider the conditional Pfaffian $L$-ensemble on $\X$ with
$\Y=\X^{(1)}\cup\cdots\cup\X^{(k)}$ and the matrix $L$ which in
the block form corresponding to the splitting
$$
\{1,\dots,2n\}\cup\bigl(\X^{(1)}\bigr)'\cup\bigl(\X^{(1)}\bigr)''\cup\cdots
\bigl(\X^{(k)}\bigr)'\cup\bigl(\X^{(k)}\bigr)''
$$
has the form
$$
L=\bmatrix 0&0&0&0&0&0&\cdots&0&0&\Xi^t\\
0&\epsilon&0&0&0&0&\cdots&0&0&0\\
0&0&0&V_1&0&0&\cdots&0&0&0\\
0&0&-V_1^t&0&0&0&\cdots&0&0&0\\
0&0&0&0&0&V_2&\cdots&0&0&0\\
0&0&0&0&-V_2^t&0&\cdots&0&0&0\\
\cdots&\cdots&\cdots&\cdots&\cdots&\cdots&\cdots&\cdots&\cdots&\cdots\\
0&0&0&0&0&0&\cdots&0&V_{k-1}&0\\
0&0&0&0&0&0&\cdots&-V_{k-1}^t&0&0\\
-\Xi&0&0&0&0&0&\cdots&0&0&0
\endbmatrix
$$
Then this conditional Pfaffian $L$-ensemble is exactly the process
defined by \tht{1.6}. We want to use Proposition 1.7 and Lemma
1.5. Writing $(\J_\Y+L)$ in $2\times 2$ block form corresponding
to the splitting
$$
\{1,\dots,2n\}\cup\Bigl(\bigl(\X^{(1)}\bigr)'\cup\bigl(\X^{(1)}\bigr)''\cup\cdots
\bigl(\X^{(k)}\bigr)'\cup\bigl(\X^{(k)}\bigr)''\Bigr)
$$
and using the notation of Lemma 1.5, we obtain that the $(i,j)$-block of
$(\J+D^{-1})$ has the form
$$
\bmatrix 0 & -V_{[i,j)}\\
V_{[j,i)}^t& V_{[1,i)}^t\epsilon V_{[1,j)}
\endbmatrix
$$
This follows, for example, from the explicit computation of the terminating
series
$$
D^{-1}=(\J_\Y+L_\Y)^{-1}=-\J_\Y\Bigl(\I+L_\Y\J_\Y+(L_\Y\J_\Y)^2+\dots+
(L_\Y\J_\Y)^{2k-1}\Bigr).
$$
Further,
$$
\gathered \Cal M=-\Xi^tV_{[1,k)}\epsilon V_{[1,k)}\Xi=-N,\\
D^{-1}C=\bmatrix V_{[1,k)}\Xi,-\epsilon V_{[1,k)}\Xi,\
V_{[2,k)}\Xi,-V_{[1,2)}^t\epsilon V_{[1,k)}\Xi,\ \dots,\
\Xi,-V_{[1,k)}^t\epsilon
V_{[1,k)}\Xi\endbmatrix^t,\\
BD^{-1}=\bmatrix \Xi^t V_{[1,k)}^t, \Xi^t V_{[1,k)}^t\epsilon,\ \Xi^t
V_{[2,k)}^t, \Xi^t V_{[1,k)}^t\epsilon V_{[1,2)},\ \dots,\ \Xi^t, \Xi^t
V_{[1,k)}^t\epsilon V_{[1,k)},
\endbmatrix
\endgathered
$$
and the $(i,j)$-block of
$$
\J_Y+(\J_\Y+L)^{-1}\bigl|_{\Y\times\Y}=\J_\Y+(D^{-1}-D^{-1}C\Cal M^{-1}BD^{-1})
$$
is readily seen to be equal to \tht{1.7}. \qed
\enddemo

\head 2. Schur process
\endhead

In the next two sections we will be extensively using the theory of symmetric
functions; we refer the reader to the book \cite{M} which contains all needed
notations and definitions.

Pick a natural number $T$ and consider all sequences of partitions
(equivalently, Young diagrams) of the form
$$
\varnothing\subset \la^{(1)}\supset \mu^{(1)}\subset \la^{(2)}\supset \mu^{(2)}
\subset \dots \supset \mu^{(T-1)}\subset \la^{(T)}\supset \varnothing. \tag 2.1
$$
To any such sequence we assign the weight
$$
\Cal W(\la,\mu)=s_{\la^{(1)}}(\rho_0^+)\,s_{\la^{(1)}/\mu^{(1)}}(\rho_1^-)
s_{\la^{(2)}/\mu^{(1)}}(\rho_1^+)\,\cdots
s_{\la^{(T)}/\mu^{(T-1)}}(\rho_{T-1}^+)\, s_{\la^{(T)}}(\rho_T^-). \tag 2.2
$$

In this formula, there is one factor for any two neighboring partitions in the
sequence. All of the factors, except for the first and the last ones, are of
the form $s_{\la/\mu}(\rho)$. The $\rho$'s here are specializations of the
algebra $\Lambda$ of symmetric functions, $s_\la$'s are the Schur functions,
and $s_{\la/\mu}$'s are the skew Schur functions.

We will use the notation $l_i=\lambda_i-i$, $m_i=\mu_i-i$. Note that
$s_{\la/\mu}$ can be written as a determinant of a submatrix of the Toeplitz
matrix $[ h_{i-j}]$\,:
$$
s_{\la/\mu}=\det[h_{l_i-m_j}]_{i,j=1}^N, \qquad N\ge
\max\{l(\la),l(\mu)\}. \tag 2.3
$$
Here $h_i$'s are the complete homogeneous symmetric functions, and $h_i=0$ if
$i<0$. Their generating function will be denoted by
$$
\sum_{k\ge 0}h_k(\rho)\,z^k=H(\rho;z).
$$

We will use the notation
$$
H(\rho';\rho'')=\sum_{\la}s_\la(\rho')s_\la(\rho'').
$$
If $\rho'$ and $\rho''$ are specializations into sets of variables
$x,y$ then one has the {\it Cauchy identity}
$$
H(x;y)=\prod_{i,j}(1-x_iy_j)^{-1}.
$$
Both sides of this identity should be viewed as formal series with elements
from $\Lambda\otimes\Lambda$; these series ``converge'' in the sense that there
are only finitely many terms of any fixed degree. In what follows we will
usually omit comments of the same kind.

For two specializations $\rho'$ and $\rho''$ we denote by $\rho'\cup\rho''$ the
specialization which adds the power sums:
$$
p_k(\rho'\cup\rho'')=p_k(\rho')+p_k(\rho''),\qquad k\ge 1.
$$
\proclaim{Proposition 2.1} The sum of the weights \tht{2.2} over all sequences
\tht{2.1} is equal to
$$
Z(\rho)=\prod_{0\le i<j\le T}H(\rho_i^+;\rho_j^-). \tag 2.4
$$
\endproclaim
\demo{Proof} Follows from the well known identity, see \cite{M, I.5.26},
$$
\sum_\nu s_{\nu/\lambda}(x) s_{\nu/\mu}(y)=H(x;y)\sum_\tau s_{\mu/\tau}(x)
s_{\la/\tau}(y).
$$
Using this formula to sum \tht{2.2} over all $\la^{(i)}$ reduces the statement
to a similar one with smaller length $T$ of the sequence \tht{2.1}. Induction
on $T$ completes the proof.\qed
\enddemo

We now consider a (formal) random point process on $\{1,\dots,T\}\times\Z$ by
assigning to a sequence \tht{2.1} the point configuration
$$
\Cal L(\la)=\left\{(1,\lambda^{(1)}_i-i)\right\}_{i\ge
1}\cup\cdots\cup \left\{(T,\lambda^{(T)}_i-i)\right\}_{i\ge 1}.
\tag 2.5
$$
The ``probability'' of this point configuration is given by the weight
\tht{2.2} divided by $Z(\rho)$. The correlation functions of this point process
are given by the following statement.

\proclaim{Theorem 2.2 (Okounkov-Reshetikhin \cite{OR})} The random
point process defined above is determinantal. In other words, for
any pairwise distinct points $(i_s,u_s)$, $1\le s\le S$, of
$\{1,\dots,T\}\times \Z$ we have the following formal series
identity
$$
\sum_{\{(i_1,u_1),\dots,(i_S,u_S)\}\subset \Cal L(\la)} \Cal
W(\la,\mu)=Z(\rho)\cdot\det_{1\le s,t\le S}\bigl[K(i_s,u_s;i_t,u_t)\bigr], \tag
2.6
$$
where
$$
K(i,u;j,v)=\frac 1{(2\pi i)^2} \oint\oint
\frac{H(\rho_{[i,T]}^{-};z)H(\rho_{[0,j)}^+;w) }
{(zw-1)H(\rho_{[0,i)}^+;z^{-1})H(\rho_{[j,T]}^-;w^{-1})}\, \frac{dzdw}{z^{u+1}
w^{v+1}}. \tag 2.7
$$
The contours for $z$ and $w$ go around 0 in the positive direction so that for
$i\le j$ we take $|z|>1$, $|w|>1$ meaning that we may expand
$$
(zw-1)^{-1}=(zw)^{-1}+(zw)^{-2}+\dots
$$
to evaluate the kernel, while for $i\ge j$ we take $|z|<1$, $|w|<1$ thus
allowing the expansion
$$
(zw-1)^{-1}=-(1+zw+(zw)^2+\dots).
$$
\endproclaim
\example{Remark 2.3} As will be shown in the proof,
\tht{2.6}-\tht{2.7} becomes a numeric equality for arbitrary
finite dimensional specializations $\rho^\pm$ with values of the
variables taken from the open unit disc, and contours in \tht{2.7}
taken close enough to the unit circle. By a simple approximation
argument it follows that \tht{2.6}-\tht{2.7} holds for arbitrary
specializations $\rho^\pm$ such that the radii of convergence of
$H(\rho^\pm_i;z)$ are strictly greater than 1, and the contours
are chosen close enough to the unit circle. As was shown by
Johansson \cite{J2}, these analytic restrictions can be further
relaxed.
\endexample

\demo{Proof} It suffices to prove \tht{2.6} when $\rho_0^+$ and $\rho_T^{-}$
are specializations into finitely many variables:
$$
\rho_0^+=(x_1,\dots,x_p),\qquad \rho_T^{-}=(y_1,\dots,y_p).
$$
If we sum \tht{2.2} over all $\mu^{(i)}$'s with $\lambda^{(j)}$'s
fixed, use \tht{2.3} and the definition of the Schur polynomial as
a ratio of two determinants, see \cite{M,I.3(3.1)}, we obtain
$$
\gathered \frac{\prod_{i=1}^p(x_iy_i)^p}{\prod\limits_{1\le i<j\le
p}(x_i-x_j)(y_i-y_j)}\cdot \det_{1\le i,j\le p}\Bigl[x_i^{l_j^{(1)}}\Bigr]
\det_{1\le i,j\le N}
W_1(l_i^{(1)},l_j^{(2)})\cdots\\\qquad\qquad\qquad\qquad\qquad \cdots
\det_{1\le i,j\le N} W_{T-1}(l_i^{(T-1)},l_j^{(T)}) \det_{1\le i,j\le
p}\Bigl[y_i^{l_j^{(T)}}\Bigr]
\endgathered
\tag 2.8
$$
where $N$ is large enough, $N\ge \max\{l(\la^{(i)})\}$, and $\Vert
W_i(x,y)\Vert_{x,y\in\Z}$ are Toeplitz matrices with symbols
$$
\sum_{m\in\Z} W_i(x+m,x)z^m=H(\rho_i^-;z)H(\rho_i^+;z^{-1}).
$$
The formula \tht{2.8} is very similar to \tht{1.2}. There are two important
differences though: the intermediate determinants in \tht{2.8} may be of any
finite size $N$, and the variables $l_j^{(i)}$ may vary over the infinite set
of all integers, not over some finite set $\X$.

However, if we are interested only in the terms of \tht{2.2} of a
small enough degree, we may restrict our attention to Young
diagrams $\la^{(i)}$ with bounded lengths of the first row and
column, which translates into boundedness of $l(\la^{(i)})$ and
$l_j^{(i)}$. Thus, in order to correctly evaluate the terms of
\tht{2.2} of a fixed degree we may choose $p$ large enough and
assume that in \tht{2.8}, $N=p$ and $l_j^{(i)}$'s vary in a finite
set. Therefore, we are in a position to apply Theorem 1.4.

The hard part in the application of Theorem 1.4 is the computation of $M^{-1}$.
Thanks to \tht{1.3} and \tht{2.4}, we know that up to terms of high degree
$$
\multline \frac {\prod_{i=1}^p(x_iy_i)^p}{\prod\limits_{1\le i<j\le
p}(x_i-x_j)(y_i-y_j)}\,\det M=\prod_{0\le i<j\le
T}H(\rho_i^+;\rho_j^-)\\=\prod_{i=1}^p H(\rho^-_{[1,T-1]};x_i)
H(\rho^+_{[1,T-1]};y_i)\cdot \prod_{i,j=1}^p\frac{1}{1-x_iy_j}\cdot \prod_{1\le
i<j\le T-1}H(\rho_i^+;\rho_j^-),
\endmultline
$$
where we use the notation
$\rho^\pm_{[i,j]}=\rho^\pm_i\cup\rho^\pm_{i+1}\cup\cdots\cup\rho^\pm_j$.

On the other hand, it is not hard to see that computing the determinant of $M$
with $k$th row and $l$th column removed is, up to terms of high degree,
equivalent to repeating the above computation with variables $x_k$ and $y_l$
removed from the specializations $\rho_0^+$ and $\rho_T^-$:
$$
\multline \frac {(x_1\cdots\hat x_k\cdots x_p\, y_1\cdots \hat y_l\cdots
y_p)^{p}}{\prod\limits_{1\le i<j\le p,\ i\ne k,j\ne l}(x_i-x_j)(y_i-y_j)}\,\det
M\binom{1\cdots \hat{k}\cdots p}{1\cdots\hat{l}\cdots p}\\=\prod_{i=1}^p
H(\rho^-_{[1,T-1]};x_i) H(\rho^+_{[1,T-1]};y_i)\cdot
\prod_{i,j=1}^p\frac{1}{1-x_iy_j}\cdot \prod_{1\le
i<j\le T-1}H(\rho_i^+;\rho_j^-)\\
\times \frac{\prod_{i=1}^p(1-x_ky_i)(1-x_iy_l)}{H(\rho^-_{[1,T-1]};x_k)
H(\rho^+_{[1,T-1]};y_l)(1-x_ky_l)}\,.
\endmultline
$$
The conclusion is that up to terms of high degree,
$$
\multline (M^{-1})_{lk}=\frac{(-1)^{k+l}\det M\binom{1\cdots \hat{k}\cdots
p}{1\cdots\hat{l}\cdots p}}{\det M}\\=\frac{x_ky_l}{\prod_{i\ne
k}(1-x_i/x_k)\prod_{j\ne
l}(1-y_j/y_l)}\,\,\frac{\prod_{i=1}^p(1-x_ky_i)(1-x_iy_l)}{H(\rho^-_{[1,T-1]};x_k)
H(\rho^+_{[1,T-1]};y_l)(1-x_ky_l)}.
\endmultline
$$
Hence, in the notation of \tht{1.4} we have
$$
\multline (\Psi M^{-1} \Phi)_{uv}\\=\sum_{k,l=1}^p
\frac{x_k^{v+1}y_l^{u+1}}{\prod_{i\ne k}(1-x_i/x_k)\prod_{j\ne
l}(1-y_j/y_l)}
\frac{\prod_{i=1}^p(1-x_ky_i)(1-x_iy_l)}{H(\rho^-_{[1,T-1]};x_k)
H(\rho^+_{[1,T-1]};y_l)(1-x_ky_l)}\\= \frac 1{(2\pi i)^2}
\oint\oint \frac{H(\rho_0^+;z^{-1})H(\rho_T^{-};w^{-1})\,z^vw^u }
{(1-zw)H(\rho_{[1,T]}^-;z)H(\rho_{[0,T)}^+;w)}\, dzdw.
\endmultline
$$
The last equality is just a formal evaluation of residues of the
integrand at the points $z=x_k$, $w=y_l$; $k,l=1,\dots,p$. Then,
using the same rule of evaluating the integrals, up to terms of
high degree, we obtain
$$
(W_{[i,k)}\Psi M^{-1}\Phi W_{[i,j)})_{uv}=\frac 1{(2\pi i)^2}
\oint\oint
\frac{H(\rho_{[0,j)}^+;z^{-1})H(\rho_{[i,T]}^{-};w^{-1})\,z^vw^u }
{(1-zw)H(\rho_{[j,T]}^-;z)H(\rho_{[0,i)}^+;w)}\, dzdw.
$$
Finally, if for $i<j$ we evaluate the residue of the right-hand side at
$w=z^{-1}$, we get
$$
-\frac{1}{2\pi i} \oint H(\rho_{[i,j)}^-;z)
H(\rho_{[i,j)}^+;z^{-1}) z^{v-u-1}dz=(-W_{[i,j)})_{uv}.
$$

Thus, \tht{1.4} implies the statement of the theorem\footnote{with
the change $(z,w)\to (w^{-1},z^{-1})$ of the integration
variables} for finite-dimensional specializations
$\rho_0^+=(x_1,\dots,x_p)$, $\rho_T^-=(y_1,\dots,y_p)$, with the
following (formal) rule of evaluating the double contour integral:
for $i\le j$ we sum up all the residues at $z=x_k$, $w=y_l$, and
for $i>j$ we also add the residue at $w=z^{-1}$.

If we now assume that all our specializations $\rho^\pm_i$ are
finite-dimensional with numeric values of the variables taken from the open
unit disc, then this evaluation rule will give the actual value of the integral
if for $i\le j$ we take the contours to be circles $|z|=|w|=1-\varepsilon$ with
small enough $\varepsilon>0$, and for $i>j$ we take the circles
$|z|=|w|=1+\varepsilon$ with small enough $\varepsilon>0$. Thus, in this case
we can evaluate the integral in a different way, by expanding $(1-zw)^{-1}$ and
all the $H$'s into Taylor series and computing the residue at $z=0$, $w=0$.
This proves our theorem for any finite dimensional specializations, and hence
for any specializations.\qed
\enddemo

\head 3. Pfaffian Schur process
\endhead

Once again, we consider sequences of Young diagrams of the form \tht{2.1}, but
the weight \tht{2.2} is replaced by
$$
\Cal
V(\la,\mu)=\tau_{\la^{(1)}}(\rho_0^+)\,s_{\la^{(1)}/\mu^{(1)}}(\rho_1^-)
s_{\la^{(2)}/\mu^{(1)}}(\rho_1^+)\,\cdots
s_{\la^{(n)}/\mu^{(n-1)}}(\rho_{n-1}^+)\, s_{\la^{(n)}}(\rho_n^-)
\tag 3.1
$$
where the symmetric functions $\tau_\la$ are defined by
$$
\tau_\la=\sum_{\kappa'\text{  is even}} s_{\la/\kappa}.
$$
\proclaim{Lemma 3.1} The symmetric function $\tau_\la$ can be
written as a Pfaffian of a Toeplitz matrix made of complete
homogeneous symmetric functions as follows:
$$
\tau_\la=\pf\left[\sum_{a\in\Z}
\left(h_{l_i-a-1}h_{l_j-a}-h_{l_i-a}h_{l_j-a-1}\right)\right]_{1\le
i,j\le 2N},\qquad l(\la)\le 2N. \tag 3.2
$$
\endproclaim
\demo{Proof} It is not hard to see that the indicator function for
partitions $\kappa$ with even conjugate and $l(\mu)\le 2N$ can be
expressed as a Pfaffian:
$$
\chi(\kappa)=\pf_{1\le i,j\le
2N}\left[\delta_{\ka_i-i-1,\,\ka_j-j}-\delta_{\ka_i-i,\,\ka_j-j-1} \right].
$$
Using the Pfaffian variant of the Cauchy-Binet formula and the
notation $k_i=\ka_i-i$, we obtain (all determinants/Pfaffians are
of size $2N\ge l(\la)$)
$$
\multline \tau_\la=\sum_\ka\det
[h_{l_i-k_j}]\pf\left[\delta_{\ka_i-i-1,\,\ka_j-j}-
\delta_{\ka_i-i,\,\ka_j-j-1}\right]\\=\sum_\ka\pf\Bigl[\Vert
h_{l_i-k_j}\Vert\cdot\Vert\delta_{\ka_i-i-1,\,\ka_j-j}-
\delta_{\ka_i-i,\,\ka_j-j-1}\Vert\cdot \Vert h_{l_i-k_j}\Vert^t\Bigr]\\=
\pf\left[\sum_{a\in\Z}
\left(h_{l_i-a-1}h_{l_j-a}-h_{l_i-a}h_{l_j-a-1}\right)\right].\qed
\endmultline
$$
\enddemo

The definition of $\tau_\la$ implies that if we specialize $\tau_\la$ into one
nonzero variable $\alpha$ then $\tau_\la(\alpha)=\alpha^{\sum_{i\ge
1}\la_{2i-1}-\la_{2i}}$ (there is a unique choice of $\kappa$ that gives a
nonzero contribution). In particular, $\tau_\la(1)=1$.

Note also that the symbol of the Toeplitz matrix in \tht{3.2}
 is equal to $$(z^{-1}-z)H(\rho;z)H(\rho;z^{-1}).$$

In addition to the notation $H(\rho';\rho'')$ introduced in the previous
section, we define
$$
H^o(\rho)=\sum_{\la'\text{  is even}}s_\la(\rho).
$$

If $\rho$ is the specialization into a set of variables $x$ then
$$
H^o(x)=\prod_{i<j}(1-x_ix_j)^{-1}.
$$

We have the following analog of Proposition 2.1.

\proclaim{Proposition 3.2} The sum of weights \tht{3.1} over all sequences
\tht{2.1} is equal to
$$
Z^o(\rho)=H^o(\rho_{[1,T]}^-)\prod_{0\le i<j\le T}H(\rho_i^+;\rho_j^-). \tag
3.3
$$
\endproclaim
\demo{Proof} As in the proof of Proposition 2.1, we sum over all $\la^{(i)}$
using the identity used there together with, see \cite{M, I.5.27},
$$
\sum_{\nu'\text{  even}}s_{\nu/\la}(x)=H^o(x)\sum_{\kappa'\text{
even}}s_{\la/\kappa}
$$
thus reducing the statement to the case of smaller $T$. \qed
\enddemo

Similarly to \S2, we consider the random point process on
$\{1,\dots,T\}\times\Z$ generated by the point configurations $\Cal L(\la)$,
see \tht{2.5}, and weights \tht{3.1}.

\proclaim{Theorem 3.3} The point process introduced above is
Pfaffian. In other words, for any pairwise distinct points
$(i_s,u_s)$, $1\le s\le S$, of $\{1,\dots,T\}\times \Z$ we have
the following formal series identity
$$
\sum_{\{(i_1,u_1),\dots,(i_S,u_S)\}\subset \Cal L(\la)} \Cal
V(\la,\mu)=Z^o(\rho)\cdot\pf\bigl[K(i_s,u_s;i_t,u_t)\bigr]_{1\le
s,t\le S}
$$
where $K(i,u;j,v)$ is a $2\times 2$ matrix kernel
$$
K(i,u;j,v)=\bmatrix
K_{11}(i,u;j,v)&K_{12}(i,u;j,v)\\K_{21}(i,u;j,v)&K_{22}(i,u;j,v)\endbmatrix
$$
whose blocks are given by:
$$
\multline K_{11}(i,u;j,v)=\frac 1{(2\pi i)^2}\\ \times \iint
\frac{(z-w)}{(z^2-1)(w^2-1)(zw-1)}\frac{H(\rho^-_{[i,T]};z)H(\rho^-_{[j,T]};w)}
{H(\rho^-_{[1,T]}\cup\rho^+_{[0,i)};z^{-1})H(\rho^-_{[1,T]}\cup
\rho_{[0,j)}^+;w^{-1})}\,\frac{dzdw}{z^uw^v}
\endmultline
$$
The integrals are taken along closed contours which go around zero in the
positive direction, and such that $|z|>1$, $|w|>1$,\footnote{This condition
means that we may use the expansions
$$
(z^2-1)^{-1}=\sum_{k\ge 0} z^{-2k-2}, \quad (w^2-1)^{-1}=\sum_{k\ge 0}
w^{-2k-2}, \quad (zw-1)^{-1}=\sum_{k\ge 0} (zw)^{-k-1}
$$ to see that this integral is a formal series of symmetric
functions. Similar comments apply to other integral below.}
$$
\multline K_{12}(i,u;j,v)=-K_{21}(j,v;i,u)\\ =\frac 1{(2\pi i)^2}
\iint \frac{(z-w)}{(z^2-1)(zw-1)w}\,
\frac{H(\rho^-_{[i,T]};z)H(\rho^-_{[1,T]}\cup \rho_{[0,j)}^+;w)}
{H(\rho^-_{[1,T]}\cup\rho^+_{[0,i)};z^{-1})H(\rho^-_{[j,T]};w^{-1})}\,
\frac{dzdw}{z^uw^v}
\endmultline
$$
The integrals are taken along closed contours which go around zero in the
positive direction, and such that $|z|>1$ and

 $\bullet$\quad  if $i\ge j$ then $|zw|>1$;

 $\bullet$\quad  if $i< j$ then $|zw|<1$.

Finally,
$$
\multline K_{22}(i,u;j,v)=\\=\frac 1{(2\pi i)^2}\iint
\frac{z-w}{zw(1-zw)}\,\frac{H(\rho^-_{[1,T]}\cup\rho^+_{[0,i)};z)
H(\rho^-_{[1,T]}\cup\rho^+_{[0,j)};w)}{H(\rho^-_{[i,T]};z^{-1})
H(\rho^-_{[j,T]};w^{-1})}\,\frac{dzdw}{z^uw^v}
\endmultline
$$
The integrals are taken along closed contours which go around zero in the
positive direction, and such that $|zw|<1$.
\endproclaim

\example{Remark 3.4} Similarly to the determinantal case of \S2, the statement
of Theorem 3.3 becomes a numeric equality if all the specializations are such
that the radii of convergence of $H(\rho_i^\pm;z)$ are strictly greater than 1
and the contours are chosen close enough to the unit circle.
\endexample

\demo{Proof} Since the computations are very similar to those in the proof of
Theorem 2.2, we will omit the necessary justifications and just produce the
formulas.

Using the similarity of \tht{3.1} and \tht{1.6}, we will compute the
correlation kernel via Theorem 1.9. Let us take $\rho_T^{-}$ to be the finite
dimensional specialization into variables $x_1,\dots,x_{2p}$. The the matrix
$N^{-1}$ is computed using \tht{3.3} in the same way as $M^{-1}$ in the proof
of Theorem 2.2 was computed using \tht{2.4}. Namely, up to terms of high
degree,
$$
\multline \frac{(x_1\cdots x_{2p})^{2p}}{\prod_{1\le i<j\le 2p}(x_i-x_j)}\,\pf
N=H^o(\rho_{[1,T]}^-)\prod_{0\le i<j\le T}H(\rho_i^+;\rho_j^-)
\\=\prod_{1\le i<j\le 2p}\frac 1{1-x_ix_j}\prod_{i=1}^{2p}H(\rho^-_{[1,T)}\cup
\rho^+_{[0,T)};x_i)  \cdot H^o(\rho_{[1,T)}^-)\prod_{0\le i<j\le
T-1}H(\rho_i^+;\rho_j^-)
\endmultline
$$
Furthermore, for $k<l$, up to terms of high degree we have
$$
\multline \frac{(x_1\cdots\hat x_k\cdots\hat x_l\cdots
x_{2p})^{2p}}{\prod_{1\le i<j\le 2p,\, i,j\ne k,l}(x_i-x_j)}\,\pf
N\binom{1\cdots \hat k\cdots \hat l\cdots 2p} {1\cdots \hat
k\cdots \hat l\cdots 2p}\\=\prod_{1\le i<j\le 2p}\frac
1{1-x_ix_j}\prod_{i=1}^{2p}H(\rho^-_{[1,T)}\cup\rho^+_{[0,T)};x_i)
\cdot
H^o(\rho_{[1,T)}^-)\prod_{0\le i<j\le T-1}H(\rho_i^+;\rho_j^-)\\
\times
\frac{\prod_{i=1}^{2p}(1-x_ix_k)(1-x_ix_l)}{(1-x_k^2)(1-x_l^2)(1-x_kx_l)
\,H(\rho^-_{[1,T)}\cup\rho^+_{[0,T)};x_k,x_l)}
\endmultline
$$
and
$$
\multline (N^{-1})_{kl}=(-1)^{k+l}\,\frac{\pf N\binom{1\cdots \hat k\cdots \hat
l\cdots 2p} {1\cdots \hat k\cdots \hat l\cdots 2p}}{\pf N}
=\frac{(x_l-x_k)x_kx_l}{\prod_{i\ne k}(1-x_i/x_k)\prod_{j\ne l}(1-x_j/x_l)}\\
\times\frac{\prod_{i=1}^{2p}(1-x_ix_k)(1-x_ix_l)}{(1-x_k^2)(1-x_l^2)(1-x_kx_l)
\,H(\rho^-_{[1,T)}\cup\rho^+_{[0,T)};x_k,x_l)}\,.
\endmultline
$$
Hence,
$$
\multline
(\Xi N^{-1}\Xi^t)_{uv}=\sum_{k,l=1}^{2p}\frac{(x_l-x_k)x_k^{u+1}x_l^{v+1}}{\prod_{i\ne k}(1-x_i/x_k)\prod_{j\ne l}(1-x_j/x_l)}\\
\times\frac{\prod_{i=1}^{2p}(1-x_ix_k)(1-x_ix_l)}{(1-x_k^2)(1-x_l^2)(1-x_kx_l)
\,H(\rho^-_{[1,T)}\cup\rho^+_{[0,T)};x_k,x_l)}\\
=\frac 1{(2\pi
i)^2}\oint\oint\frac{(w-z)H(\rho_T^-;z^{-1},w^{-1})z^uw^v
}{(1-z^2)(1-w^2)(1-zw)H(\rho^-_{[1,T)}\cup\rho^+_{[0,T)};z,w)}\,dzdw
\endmultline
$$
The integral is understood as the sum of residues at the points
$z,w=x_1,\dots,x_{2p}$. Taking convolutions of this expression with $V_i$'s,
which are Toeplitz matrices with symbols $H(\rho_i^-;z)H(\rho_i^+;z^{-1})$, and
with $\epsilon$ which is also Toeplitz with symbol
$(z^{-1}-z)H(\rho_0^+;z)H(\rho_0^+;z^{-1})$, we obtain, in the notation of
\tht{1.7},
$$
\multline (V_{[i,T)}\Xi N^{-1}\Xi^t V^t_{[j,T)})_{uv}\\=\frac
1{(2\pi i)^2} \iint \frac{(w-z)z^uw^v}{(1-z^2)(1-w^2)(1-zw)}
\frac{H(\rho^-_{[i,T]};z^{-1})H(\rho^-_{[j,T]};w^{-1})}
{H(\rho^-_{[1,T]}\cup\rho^+_{[0,i)};z)H(\rho^-_{[1,T]}\cup
\rho_{[0,j)}^+;w)}\,{dzdw}.
\endmultline
$$
Inverting the variables of integration yields the expression for $K_{11}$.

Furthermore,
$$
\multline (V_{[i,T)}\Xi N^{-1}\Xi^t V^t_{[1,T)}\epsilon
V_{[1,j)})_{uv}\\=\frac 1{(2\pi i)^2} \iint
\frac{(w-z)z^uw^{v}}{(1-z^2)w(1-zw)}
\frac{H(\rho^-_{[i,T]};z^{-1})H(\rho^-_{[1,T]}\cup
\rho_{[0,j)}^+;w^{-1})}
{H(\rho^-_{[1,T]}\cup\rho^+_{[0,i)};z)H(\rho^-_{[j,T]};w)}\,{dzdw}.
\endmultline
$$
Note that the residue of this integral at $w=z^{-1}$ equals $(i<j)$
$$
-\frac 1{2\pi i}\int
{H(\rho^-_{[i,j)};z^{-1})H(\rho_{[i,j)}^+;z)}z^{u-v-1}dz=(-V_{[i,j)})_{uv},
$$
which is the second term in the (1,2)-entry of \tht{1.7}. This proves the
formula for $K_{21}$ and $K_{12}$.

Finally,
$$
\multline (-V^t_{[1,i)}\epsilon V_{[1,T)}\Xi N^{-1}\Xi^t V^t_{[1,T)}\epsilon
V_{[1,j)})_{uv}\\=\frac 1{(2\pi i)^2} \iint
\frac{(w-z)z^uw^{v}}{zw(1-zw)}\,\frac{H(\rho^-_{[1,T]}\cup\rho^+_{[0,i)};z^{-1})
H(\rho^-_{[1,T]}\cup\rho^+_{[0,j)};w^{-1})}{H(\rho^-_{[i,T]};z)
H(\rho^-_{[j,T]};w)}\,dzdw,
\endmultline
$$
and the residue of the integral at $w=z^{-1}$ gives
$$
\frac{1}{2\pi i}\int (z-z^{-1})H(\rho_{[1,j)}^-\cup\rho_{[0,i)}^+;z^{-1})
H(\rho_{[1,i)}^-\cup\rho_{[0,j)}^+;z)z^{u-v-1}dz=(V^t_{[1,i)}\epsilon
V_{[1,j)})_{uv}
$$
as is needed in the (2,2)-block of \tht{1.7}. \qed
\enddemo

\head{4. Abstract processes}
\endhead

Given a conditional determinantal or Pfaffian $L$-ensemble, the
associated collection of minors determines a point in
$\R^{2^n}=(\R^2)^{\otimes n}$. For algebraic purposes, it is nicer
to consider the complex analogue of this notion; we thus obtain
the following definitions.

\example{Definition 4.1} Let $n$ be a positive integer.  A nonzero
point $p\in (\C^2)^{\otimes n}$ is {\it determinantal} if there
exists an integer $m\ge 0$ and an $(n+m)\times (n+m)$ matrix $K$
such that for $S\subset \{1,2,\dots n\}$,
$$
p_S = \det_{S\cup \{n+1,\dots,n+m\}}(K).
$$
The point is {\it Pfaffian} if there exists a $2\times 2$ matrix
valued $(n+m)\times (n+m)$ skew-symmetric matrix $K$ such that
$$
p_S = \pf_{S\cup \{n+1,\dots,n+m\}}(K)
$$
for all $S\subset \{1,2,\dots n\}$.

A point process whose correlation functions $\rho(S)$ are given by
minors $p_S$ as above is called {\it conditional determinantal}
(or {\it conditional Pfaffian\/}).
\endexample

 If we replace $K$ by the block matrix
$$
\bmatrix
K & 0\\
0 & s
\endbmatrix,
\text{ respectively } \bmatrix
K & 0\\
0 & \bmatrix 0&s\\-s&0\endbmatrix
\endbmatrix,
$$
for some nonzero scalar $s$, this simply multiplies $p_S$ by $s$,
and thus the conditions depend only on the corresponding points in
the projective space $\P^{2^n-1}(\C)$.

Now, adding a multiple of one of the last $m$ rows/columns of $K$
to one of the first $n$ rows/columns of $K$ leaves $p$ unchanged,
as does an arbitrary change of basis applied to the last $m$
rows/columns.  It follows that we may choose $K$ of the form
$$
\bmatrix
A & B & 0\\
C & 0 & 0\\ 0 & 0 & D
\endbmatrix,
$$
where $A$ is $n\times n$ and $D$ is diagonal and invertible.  But
this gives the same point, projectively, as
$$
\bmatrix
A & B\\
C & 0
\endbmatrix.
$$
It follows that in the definition of determinantal or Pfaffian
points, it suffices to consider $m=n+1$.  In particular, the set
of such points is an algebraic set, as the image of the space of
matrices under a polynomial map.

\proclaim{Theorem 4.2} The set of determinantal (resp. Pfaffian)
points in $\P^{2^n-1}(\C)$ is invariant under the natural action
of the group $\GL_2(\C)^n\ltimes S_n$.
\endproclaim

Here the $j$th copy of $\GL_2(\C)$ acts on $p$ by
$$\align
\left(\bmatrix a&b\\c&d\endbmatrix_j p\right)_S &= ap_S+bp_{S\cup\{j\}}\\
\left(\bmatrix a&b\\c&d\endbmatrix_j p\right)_{S\cup j} &=
    cp_S+dp_{S\cup\{j\}},
\endalign
$$
and $S_n$ acts in the obvious way (permuting the tensor factors).

\demo{Proof} We consider the determinantal case; the Pfaffian case
is analogous.

The invariance under $S_n$ is immediate, and thus invariance under
the full group will follow from invariance under the first copy of
$\GL_2(\C)$.

Suppose $p$ is determinantal, with kernel
$$
K = \bmatrix
a & \vec{b} & \vec{c}\\
\vec{d} & E & F\\
\vec{g} & H & M
\endbmatrix.
$$
Multiplying the first row or first column by $\alpha$ replaces $p$
by
$$
\bmatrix 1 & 0 \\ 0 & \alpha \endbmatrix_1 p\,,
$$
and thus the latter is determinantal.  Similarly, replacing $a$ by
$a+a_0$ takes $p$ to
$$
\bmatrix 1 & 0 \\ a_0 & 1\endbmatrix_1 p\,.
$$
We thus have invariance under a Borel subgroup of $\GL_2(\C)$; it
will thus suffice to consider the corresponding Weyl group.  In
other words, we need to show invariance under
$$
\bmatrix 0&1\\1&0\endbmatrix_1;
$$
in particular, that determinantal/Pfaffian processes are closed
under taking symmetric differences.\footnote{The observation that
the set of determinantal processes is invariant under taking
symmetric differences is due to Kerov, cf. \cite{BOO, A.3}.} In
fact, $\left(\bmatrix 0&1\\1&0\endbmatrix_1p\right)$ can be
obtained from the $(n+m+1)\times (n+m+1)$ kernel
$$
K' = \bmatrix
 0 &    0    &   0     &    1   \\
 0 &    E    &   F     & \vec{d}\\
 0 &    H    &   M     & \vec{g}\\
-1 &  \vec{b}& \vec{c} &  a
 \endbmatrix.
$$
The invariance claim follows.
\enddemo

\example{Remark 4.3} Note in particular that for any probability
distribution $p$,
$$
\left(\bmatrix 1 & 1 \\ 0 & 1\endbmatrix^{\otimes n}p \right)_S =
\sum_{T\supset S} p_S,
$$
which should be viewed as the correlation function of $p$.  We
thus arrive at an {\it a priori} nonobvious conclusion that every
conditional determinantal (Pfaffian) process is a conditional
determinantal (Pfaffian) $L$-ensemble and {\it vice versa}. (Note
that the converse statement also follows from Proposition 1.2.)
\endexample

Let $D_n$ be the topological closure of the set of determinantal
points, and let $P_n$ be the closure of the set of Pfaffian
points; of course both of these are projective varieties.  Now,
the generic point of either set satisfies $p_\emptyset\ne 0$, and
thus can be obtained from an $n\times n$ kernel.  Thus
na\"{\i}vely, we should have the dimensions
$$
\dim(D_n) \sim n^2-n+1,\qquad \dim(P_n)\sim n(2n-1)-3n,
$$
in each case the difference of the dimension of the space of
kernels and the dimension of the set of ``equivalent'' kernels.
(For $D_n$, this entails conjugation by diagonal matrices, while
for $P_n$, it entails the natural action of $\SL_2(\C)^n$ on the
kernel.)  Of course, if there exist inequivalent kernels for the
same point, or if the generic kernel has an automorphism, these
formulas fail, but this happens only for small $n$. We in fact
have the following.

\proclaim{Theorem 4.4} For all $n$, $\dim(D_n)=n^2-n+1$; in
particular, for $n\le 3$, $D_n=\P^{2^n-1}(\C)$.  Similarly, for
$n\ge 5$, $\dim(P_n)=2n(n-2)$, while for $n\le 4$,
$P_n=\P^{2^n-1}(\C)$.
\endproclaim

\demo{Proof} For $D_n$, the generic $n\times n$ kernel has a
canonical form (in which the off-diagonal entries of the first row
are all 1), from which we readily determine that it has no
automorphisms, and is uniquely determined by the associated point
(in fact by the coordinates of that point on sets of size $\le
2$).  Thus the na\"{\i}ve dimension count is in fact accurate.

For $P_n$, both properties fail for small $n$.  For $n=1$, every
kernel is invariant under $\SL_2(\C)$, while for $n=2$, the
generic kernel can be taken to the form
$$
\bmatrix
 0 & a & 0 & -b\\
-a & 0 & b &  0\\
 0 &-b & 0 & -c\\
 b & 0 & c &  0
\endbmatrix,
$$
invariant under the diagonal subgroup of $\SL_2(\C)^2$.  For
$n=3$, the generic kernel still has a 1-dimensional automorphism
group; finally for $n\ge 4$, the generic kernel has no
automorphisms. Since for $n\le 3$, $P_n\supset
D_n=\P^{2^n-1}(\C)$, we have $\dim(P_n)=2^n-1$ for $n\le 3$, and
thus the generic automorphism group is the only correction to the
dimension formula; in particular, the generic point in $P_1$,
$P_2$, $P_3$ determines a unique kernel up to equivalence.

For $n=4$, the above dimension count is too high; it gives 16 out
of a possible 15, suggesting that the generic point determines a
one-parameter family of equivalence classes of kernels.  By direct
computation with a random Pfaffian point, one can show the
existence of a point with such a family, showing that
$\dim(P_4)\ge 15$ and thus $P_4=\P^{2^n-1}(\C)$.

Similarly, for $n=5$, it suffices to find a (random) point having
a unique kernel up to equivalence; the lack of automorphisms gives
rise to a canonical form, showing that this uniqueness extends to
all larger $n$.
\enddemo

The first nontrivial instances are thus $D_4$ and $P_5$.  The
structure of $D_4$ can be deduced from the following fact.

\proclaim{Proposition 4.5} Let $p\in \P^{2^4-1}(\C)$ be a point
such that $p_S=0$ unless $|S|=2$. Then $p$ is determinantal.
\endproclaim

\demo{Proof} Equivalently, we may assume that $p$ is supported on
the six sets
$$
\emptyset,\{1,2,3,4\},\{1,3\},\{1,4\},\{2,3\},\{2,4\};
$$
and by symmetry and rescaling assume that $p_\emptyset=1$. But
then the kernel
$$
\bmatrix
0&0&1&-p_{\{1,4\}}\\
0&0&\frac{-p_{\{2,3\}}}{x}&1\\
-p_{\{1,3\}}& x & 0&0\\
1& -p_{\{2,4\}}& 0&0
\endbmatrix
$$
works, for a suitable choice of $x$.
\enddemo

\proclaim{Theorem 4.6} A point in $\P^{2^4-1}(\C)$ is
determinantal if and only if it is in the $\GL_2(\C)^{\otimes
4}$-orbit of a point supported on sets of size 2. Equivalently,
$D_4$ is the codimension 2 variety
$\nabla_{\text{node}}(\emptyset)$, in the notation of \cite{WZ},
where we have identified $\C^{2^4}$ with the space of multilinear
polynomials on $(\P^1)^4$; in other words, $D_4$ is the variety of
multilinear polynomials with two critical points in general
position.
\endproclaim

\demo{Proof} Given a multilinear polynomial with two critical
points in general position, we may act by $\GL_2(\C)^{\otimes 4}$
to put the critical points at $(0,0,0,0)$,
$(\infty,\infty,\infty,\infty)$; but then the corresponding point
in $\P^{15}(\C)$ is determinantal by the proposition.  The
remaining claims follow by comparing dimensions.
\enddemo

Note that although this gives a fairly simple direct
characterization of $D_4$, the variety itself is fairly
complicated.  In fact, one can show that the variety has degree
28, with ideal generated by a whopping 718 degree 12 polynomials.

For $P_5$, the situation is even worse; although dimension
considerations show that $P_5$ is a hypersurface, and thus cut out
by a single $\GL_2$-invariant polynomial, experimentation over
finite fields suggests that this polynomial has degree $1146$.  We
have also been unable to find any sort of natural direct
characterization of $P_5$.

\end